# Theoretical analysis of frequency variation tolerance between elements of RTD-THz oscillator arrays for mutual locking

Masahiro Asada[1] and Safumi Suzki

Institute of Integrated Research, Institute of Science Tokyo,

Meguro-ku, Tokyo Japan.

1) E-mail: asada@pe.titech.ac.jp

**Abstract:**

Array configuration is one of the effective ways to increase the output power of terahertz oscillators using resonant tunneling diodes. Mutual locking between the array elements results in coherent single-spectrum oscillation and a narrow radiation beam. However, frequency variation between elements disturbs the mutual locking. In this paper, the frequency variation that can be tolerated for mutual locking is approximately derived for an array with arbitrary coupling configuration. Using this result, the element-number dependence of the frequency variation tolerance for 1D and 2D arrays is calculated. In 1D arrays, the frequency variation tolerance decreases inversely proportional to the square root of the element number. In 2D arrays, as the number of columns is increased while keeping the number of elements per column constant, the frequency variation tolerance increases for small number of columns, reaches a maximum at square configuration, and then decreases with increasing number of columns. The element-number dependence in 2D arrays is smaller than that in 1D arrays.

## 1. Introduction

The terahertz (THz) band, covering a frequency range from approximately 0.1 THz to 10 THz, is expected to have diverse applications, including imaging, chemical and biological analysis, and communications [1–3]. Compact solid-state THz sources are key devices for these applications. As the THz band lies between millimeter waves and light waves, various types of THz sources, including both optical [4–9] and electronic [10–23] devices, have been investigated. Among them, resonant tunneling diodes (RTDs) are promising candidates for room-temperature THz sources [20–23]. Currently, oscillations up to 1.98 THz [24] and output powers exceeding 1 mW for single devices via transmission lines or antenna radiation [25, 26] and output powers exceeding 11 mW for arrays [27] have been achieved.

Increasing the output power of THz sources is crucial for various applications. High-power RTD oscillators have been investigated with two approaches: increasing the output power of a single oscillator and array configuration of low-power oscillators. Of course, arraying high-power single oscillators will be further effective. High-power single oscillators have been achieved by large-area RTDs without frequency lowering, the impedance matching between the load and RTD, and the reduction of the resonator losses [25, 26, 28, 29]. Attempts have also been made to increase the output power by arrays using various coupling configurations and element numbers [27, 30–35].

Arrays with weak or unintentional coupling between elements exhibit complex spectra due to insufficient mutual locking caused by variations in oscillation frequencies of elements [30]. A sufficient coupling between elements results in mutual locking across all elements, resulting in coherent single-spectrum oscillation [27,31–36] as well as narrow radiation beam [27]. Frequency variations among individual elements must be minimized to obtain stable mutual locking. Although the oscillation frequency of an RTD oscillator can be slightly tuned by adjusting the bias voltage [22], it is not practical to adjust the frequency of each array element individually. It is therefore necessary to make the frequencies of the array elements as uniform as possible in the device fabrication. However, the relationship between array size and acceptable frequency variations has not been fully elucidated experimentally or theoretically.

In this paper, theoretical analysis is presented on the frequency variation tolerance for mutual locking among array elements. The frequency variation that can be tolerated for mutual locking is approximately derived for an array with arbitrary coupling configuration. First, the analytical method for a two-element array is described. Next, this method is extended to a multi-element array. The frequency variation tolerance and its dependence on the number of array elements are then calculated for 1D and 2D arrays with several coupling configurations.

## 2. Analysis of a Two-Element Array

A two-element oscillator array is analyzed at first, in which two RTDs are connected via a coupling part, as shown in the schematic plan view of Fig. 1(a). The coupling part is depicted as a slot line, but it can also be a lumped constant such as resistance, capacitance, or inductance [32–35]. Figure 1(b) shows a circuit diagram of the array in which the antenna and coupling part are collectively represented as a two-terminal pair network.

The analysis is performed using the equivalent circuit shown in Fig. 1(b). The actual structure of the two-element array is not limited to that shown in Fig. 1(a), as long as it can be expressed as Fig. 1(b). The voltage and current are assumed to oscillate at a single angular frequency $\omega$, neglecting harmonic components. For example, the voltage across RTD1 is $v_1 e^{j\omega t} +$ complex conjugate (c.c.), of which only $v_1$ is shown in Fig. 1(b). $G_{RTD}$ is the absolute value of the negative differential conductance of RTD, which is assumed to be the same for both RTDs. Due to nonlinearity, $G_{RTD}$ is expressed as $G_{RTD}(v) = a - (3/4)\,b|v|^2$ ($v = v_1$ or $v_2$), where $a$ and $b$ are constants [22][23].

The voltages and currents of the RTDs are written using the $Y$ matrix of the antenna and coupling part as follows:

$$\begin{bmatrix} G_{RTD}(v_1)\,v_1 \\ G_{RTD}(v_2)\,v_2 \end{bmatrix} = \begin{bmatrix} i_1 \\ i_2 \end{bmatrix} = \begin{bmatrix} y_{11} & y_{12} \\ y_{12} & y_{22} \end{bmatrix} \begin{bmatrix} v_1 \\ v_2 \end{bmatrix}. \tag{1}$$

From eq. (1),

$$\begin{bmatrix} y_{11} - G_{RTD}(v_1) & y_{12} \\ y_{12} & y_{22} - G_{RTD}(v_2) \end{bmatrix} \begin{bmatrix} v_1 \\ v_2 \end{bmatrix} = 0. \tag{2}$$

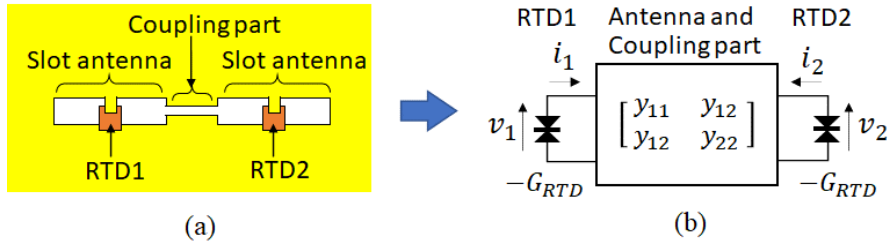

**Fig. 1** Two-element coupled oscillator array. (a) Schematic plan view and (b) expression with a two-terminal pair network.

The condition to start the oscillation (power condition) is that the real part of the eigenvalue of the coefficient matrix of eq. (2) is zero or negative [33] with $v_1$ and $v_2$ in $G_{RTD}$ set to 0. This gives

$$\mathrm{Re}\left[\left(\frac{y_{11}+y_{22}}{2}\right)\pm\sqrt{\left(\frac{y_{11}-y_{22}}{2}\right)^2+y_{12}^2}\right]\leq a\,, \tag{3}$$

The angular frequency of oscillation $\omega$ can be calculated by setting the imaginary part of the eigenvalue to zero:

$$\mathrm{Im}\left[\left(\frac{y_{11}+y_{22}}{2}\right)\pm\sqrt{\left(\frac{y_{11}-y_{22}}{2}\right)^2+y_{12}^2}\right]=0\,. \tag{4}$$

If $y_{11}=y_{22}$ $(=y)$, eqs. (3) and (4) become $\mathrm{Re}(y\pm y_{12})\leq a$ and $\mathrm{Im}(y\pm y_{12})=0$, respectively. Corresponding to the $\pm$ signs in these equations, $v_1=\pm v_2$. The $+$ sign is the even mode (in-phase mode), and the $-$ sign is the odd mode (anti-phase mode). Even if $y_{11}$ and $y_{22}$ are slightly different, the correspondence between the $\pm$ signs in eqs. (3) and (4) and the phase relationship between $v_1$ and $v_2$ still remains.

Hereafter, the following equations are assumed, to take into account that $y_{11}$ and $y_{22}$ are slightly different.

$$y_{11}=G_L+j\omega C_1+\frac{1}{j\omega L}\;,\;\;y_{22}=G_L+j\omega C_2+\frac{1}{j\omega L}\;. \tag{5}$$

The difference between $y_{11}$ and $y_{22}$ is assumed to be due only to the difference in capacitances $C_1$ and $C_2$, with all other elements remaining the same. Since the capacitances are mostly associated with the RTDs, this assumption means that the two RTDs have the same $G_{RTD}$, with only slight differences in capacitances. If the areas or layer thicknesses of RTDs differ slightly, not only the capacitances but also $G_{RTD}$ will differ, but the above assumption is sufficient for analyzing the effect of the differences in the oscillation frequencies of the two RTDs on mutual locking.

Substituting eq. (5) in eq. (2), the following equation is obtained:

$$\begin{bmatrix} G_{RTD}(v_1) - G_L + j\dfrac{1-\omega^2/\omega_1^2}{\omega L} & -y_{12} \\[2mm] -y_{12} & G_{RTD}(v_2) - G_L + j\dfrac{1-\omega^2/\omega_2^2}{\omega L} \end{bmatrix} \begin{bmatrix} v_1 \\ v_2 \end{bmatrix} = 0 \,. \tag{6}$$

where $\omega_1 = 1/\sqrt{LC_1}$ and $\omega_2 = 1/\sqrt{LC_2}$, which are the oscillation angular frequencies of RTD1 and RTD2, respectively, when $y_{12} = 0$, i.e., RTD1 and RTD2 are not coupled.

The oscillation condition (eq. (3)) becomes $G_L \pm \mathrm{Re}\left(\sqrt{y_{12}^2 - \Delta C^2/4}\right) \simeq G_L \pm \mathrm{Re}(y_{12}) \le a$ , where $\Delta C = C_1 - C_2$ , and $|\Delta C| \ll |y_{12}|$ is assumed. When $\mathrm{Re}(y_{12}) > 0$, the odd mode will oscillates, because it can satisfy the oscillation condition easier than the even mode, and when $\mathrm{Re}(y_{12}) < 0$, the even mode will oscillates. Since the total length of the antenna and the coupling part is usually close to half the wavelength, $\mathrm{Re}(y_{12})$ is likely to be positive, and thus, the odd mode easily oscillates, but it is possible to change the sign of $\mathrm{Re}(y_{12})$ by reversing the connections of the coupling part [35] or by using a long transmission line for the coupling part [27] (see Appendix 5).

Assuming that $|C_1 - C_2|$ is much smaller than $C_1$ and $C_2$, then $\omega \simeq \omega_1 \simeq \omega_2$ in eq. (6). This approximation is used in the analysis below except for the difference between the angular frequencies which affects the mutual locking between the RTDs. Substituting $v_1 = V_1 e^{j\phi_1}$ $(V_1 = |v_1|,\ \phi_1 = \angle v_1)$ and $v_2 = V_2 e^{j\phi_2}$ $(V_2 = |v_2|,\ \phi_2 = \angle v_2)$ in eq. (6) and extracting the imaginary part, the following equations are obtained.

$$\frac{\omega_1 - \omega}{\omega_1}\frac{V_1}{V_2} + \kappa_r \sin(\phi_2 - \phi_1) + \kappa_i \cos(\phi_2 - \phi_1) = 0 \,, \tag{7}$$

$$\frac{\omega_2 - \omega}{\omega_1}\frac{V_2}{V_1} + \kappa_r \sin(\phi_1 - \phi_2) + \kappa_i \cos(\phi_1 - \phi_2) = 0 \,, \tag{8}$$

where $\kappa_r$ and $\kappa_i$ are the real and imaginary parts of the coupling coefficient, respectively, given by

$$\kappa_r = -\frac{\omega_1 L}{2}\mathrm{Re}(y_{12}) \,, \quad \kappa_i = -\frac{\omega_1 L}{2}\mathrm{Im}(y_{12}) \,. \tag{9}$$

$\kappa_r$ and $\kappa_i$ correspond to $\varepsilon\kappa_r$ と $\varepsilon\kappa_i$ in ref. [36]. Since the difference between $C_1$ and $C_2$ is small, $y_{11} \simeq y_{22}$. Therefore, $V_1 \simeq V_2$, and the following equation is obtained from the difference between eqs. (7) and (8).

$$\sin(\phi_1 - \phi_2) \simeq \frac{1}{2\kappa_r}\frac{\omega_1 - \omega_2}{\omega_1} \,. \tag{10}$$

The phase difference $\phi_1 - \phi_2$ that satisfies this equation lies between 0 and $\pi$ when the right-hand side of eq. (10) is positive, and between $\pi$ and $2\pi$ when it is negative. In either case, a phase difference close to 0 or $2\pi$ corresponds to the even mode, and a phase difference close to $\pi$ corresponds to the odd mode. The phase difference can be controlled by the difference between $\omega_1$ and $\omega_2$ , or by varying $\kappa_r$ (e.g., varying the coupling distance between the RTDs) when $\omega_1 \neq \omega_2$.

Let the right-hand side of eq. (10) be $\Omega_s$. The condition for RTDs to synchronize with each other (i.e., to indicate the mutual locking) is given by $-1 \leq \Omega_s \leq 1$. Below, we will calculate the probability of mutual locking (the probability that $-1 \leq \Omega_s \leq 1$) when $\omega_1$ and $\omega_2$ vary due to errors in device fabrication.

When the variations of $\omega_1$ and $\omega_2$ are independent and follow Gaussian distributions, the variation of $\omega_1 - \omega_2$ also follows a Gaussian distribution due to the reproductive property of Gaussian distributions. Neglecting the variation of $\omega_1$ in the denominator on the right-hand side of eq. (10) compared to the variation of $\omega_1 - \omega_2$, the variation of $\Omega_s$ is expressed by the Gaussian distribution $f(\Omega_s)$ shown in Fig. 2. Assuming the mean value and squared variance of $\omega_1$ and $\omega_2$ are equal ($\langle\omega_1\rangle = \langle\omega_2\rangle = \langle\omega\rangle$ and $\langle\Delta\omega_1^2\rangle = \langle\Delta\omega_2^2\rangle = \langle\Delta\omega^2\rangle$), the mean value and squared variance of $\Omega_s$ are given by the reproductive property of Gaussian distributions as follows:

$$\langle\Omega_s\rangle = \frac{1}{2\kappa_r\omega_1}(\langle\omega_1\rangle - \langle\omega_2\rangle) = 0\ , \tag{11}$$

$$\langle\Delta\Omega_s^2\rangle = \left(\frac{1}{2\kappa_r\omega_1}\right)^2(\langle\Delta\omega_1^2\rangle + \langle\Delta\omega_2^2\rangle) = \frac{1}{2(\kappa_r\omega_1)^2}\langle\Delta\omega^2\rangle\ . \tag{12}$$

The probability $p$ that $\Omega_s$ lies within the locking range is given by

$$p = \int_{-1}^{1}\frac{\exp\left(-\frac{\Omega_s^2}{2\langle\Delta\Omega_s^2\rangle}\right)}{\sqrt{2\pi\langle\Delta\Omega_s^2\rangle}}\,d\Omega_s = \mathrm{Erf}\left(\frac{1}{\sqrt{2\langle\Delta\Omega_s^2\rangle}}\right) = \mathrm{Erf}\left(\frac{|\kappa_r|\omega_1}{\sqrt{\langle\Delta\omega^2\rangle}}\right), \tag{13}$$

where Erf is the error function ($\mathrm{Erf}(x) = (2/\sqrt{\pi})\int_0^x e^{-t^2}dt$). From eq. (13), when the relative variation of oscillation frequency ($\sqrt{\langle\Delta\omega^2\rangle}/\omega_1 = \sqrt{\langle\Delta f^2\rangle}/f_1$) due to device fabrication error is given, the probability $p$ of mutual locking can be determined. Conversely, when $p$ is specified, the relative variation of oscillation frequency allowable in device fabrication can be expressed as follows.

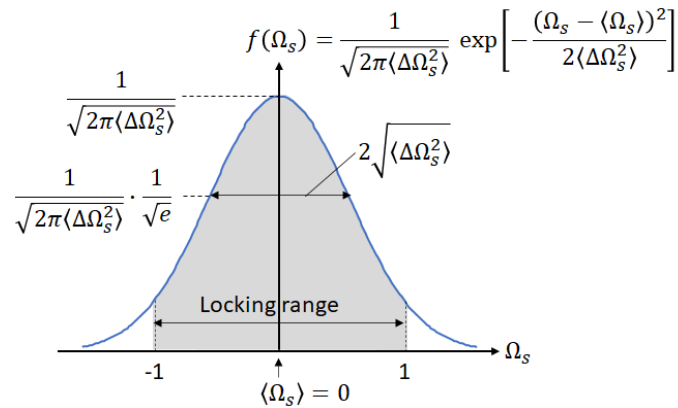

**Fig.2** Distribution of $\Omega_s$ (right-hand side of eq. (10)) due to errors in device fabrication.

$$\frac{\sqrt{\langle \Delta f^2 \rangle}}{f_1} = \sqrt{2}\, c(p)\, |\kappa_r| \ , \tag{14}$$

where the coefficient $c(p)$ $(= \sqrt{\langle \Delta \Omega_s^2 \rangle})$ is obtained from the relationship $p = \mathrm{Erf}\big(1/\sqrt{2}\,c\big)$, and is shown in Fig. 3

As a numerical example, when $\mathrm{Re}(y_{12}) \sim 2$ mS [33], $C_1 \sim 10$ fF, and $f_1 = \omega_1/2\pi \sim 1$ THz $(L = 1/\omega_1^2 C_1)$, eq. (9) gives $|\kappa_r| \sim 0.02$, and $\sqrt{\langle \Delta f^2 \rangle}/f_0 \sim 0.02$ $(\sqrt{\langle \Delta f^2 \rangle} \sim 20$ GHz$)$ for $p = 95\%$.

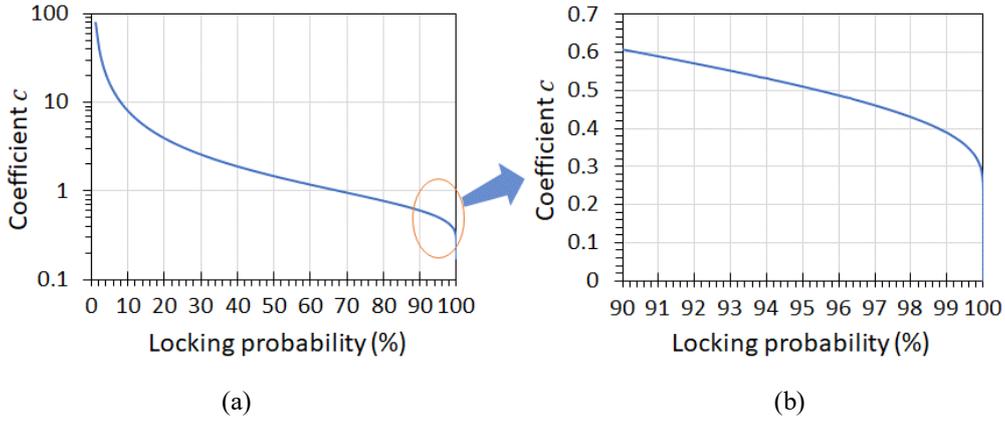

(a)                                           (b)

**Fig. 3** Relationship between the probability of mutual locking $p$ and the coefficient $c(p)$ in eq. (14). (b) is an expansion of the region 90-100% of (a).

## 3. Analysis of Multi-Element Arrays

The above analysis of the two-element array is extended to the $N$-element array shown in Fig. 4. The equation corresponding to eq. (2) is given by

$$\begin{bmatrix} y_{00} - G_{RTD}(v_0) & y_{01} & \cdots & y_{0\,N-1} \\ y_{01} & y_{11} - G_{RTD}(v_1) & \cdots & y_{1\,N-1} \\ \vdots & & \ddots & \vdots \\ y_{0N-1} & & \cdots & y_{N-1\,N-1} - G_{RTD}(v_{N-1}) \end{bmatrix} \begin{bmatrix} v_0 \\ \vdots \\ v_{N-1} \end{bmatrix} = 0. \tag{15}$$

As with the two-element case, $G_{RTD}$'s of RTDs are assumed equal while the capacitances distribute slightly. The oscillation condition is that the real part of the eigenvalue of the coefficient matrix in eq. (15) is zero or negative with $v_n$ $(n = 0,1, \cdots, N-1)$ in $G_{RTD}$ set to 0. There are $N$ eigenvalues, and their eigenvectors form the supermodes with different distributions of $v_n$'s [33]. Of these, the supermode that satisfies the oscillation condition with the smallest real part of the eigenvalue will oscillates. In many cases, the highest-order supermode easily oscillates, in which the voltages of adjacent elements have opposite phases with each other. To obtain radiation in the front direction, it is necessary to design $y_{nk}$ so that the fundamental supermode oscillates [35], or to arrange the antenna so that front radiation is obtained even with the highest-order supermode [27, 31–34].

The equations corresponding to eqs. (7) – (9) are given by

$$\frac{\omega_n - \omega}{\omega_0} + \sum_{k=0}^{N-1} \kappa_{r(nk)} \frac{V_k}{V_n} \sin(\phi_k - \phi_n)$$

$$+ \sum_{k=0}^{N-1} \kappa_{i(nk)} \frac{V_k}{V_n} \cos(\phi_k - \phi_n) = 0$$

$$(n = 0,1,\cdots,N-1) \ , \quad (16)$$

$$\left. \begin{aligned} \kappa_{r(nk)} &= -\frac{\omega_0 L}{2} \mathrm{Re}(y_{nk}) \\ \kappa_{i(nk)} &= -\frac{\omega_0 L}{2} \mathrm{Im}(y_{nk}) \end{aligned} \right\} \quad (17)$$

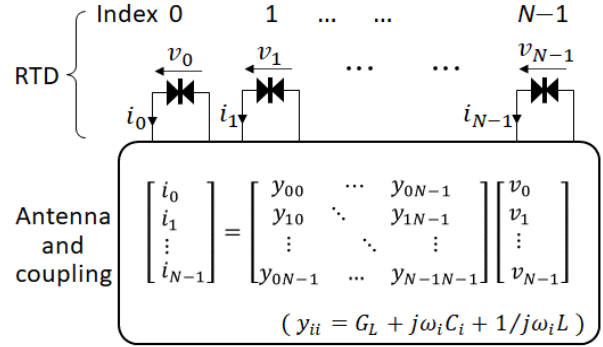

**Fig.4** *N*-element coupled oscillator array.

Since $\kappa_{r(nk)}$ and $\kappa_{i(nk)}$ represent the coupling between elements, $\kappa_{r(nk)}$ and $\kappa_{i(nk)}$ are set to zero for $k = n$ .

As in the two-element case, the voltage amplitudes $V_n$'s of the elements are assumed approximately equal hereafter. In the two-element case, if the admittances of the elements are equal, $V_n$'s are exactly equal, but in the multi-element case, even if the admittances of the elements are equal, $V_n$'s distribute according to the supermode. However, the ratio of $V_n$'s is small between the elements near the center of the array, where the phase difference due to frequency variation is large and has a significant effect on mutual locking, as described in Appendix 1 and the calculation examples below.

Under the above approximation, by summing up all of eq. (16) for $n = 0,1,\cdots,N-1$, the oscillation angular frequency is obtained as follows from $\kappa_{r(nk)} = \kappa_{r(kn)}$ and $\kappa_{i(nk)} = \kappa_{i(kn)}$ :

$$\omega \simeq \frac{1}{N} \sum_{n=0}^{N-1} \omega_n + \frac{1}{N} \omega_0 \sum_{n=0}^{N-1} \sum_{k=0}^{N-1} \kappa_{i(nk)} \cos(\phi_k - \phi_n) \ . \tag{18}$$

The oscillation frequency is thus given by the sum of the average of those of the elements and a term proportional to the imaginary part of the coupling coefficient. $\phi_k - \phi_n \simeq 0$ for the fundamental supermode, and $\phi_k - \phi_n \simeq \pm\pi$ for the highest-order supermode in the second term of eq. (18).

In the following analysis, the imaginary part of the coupling coefficient $\kappa_{i(nk)}$ is assumed much smaller than the real part $\kappa_{r(nk)}$ and is neglected. The mutual locking is less stable if $|\kappa_{i(nk)}|$ is large for a nearest-neighbor coupled array with $N > 2$ [36]. Therefore, an array structure should be designed so that the coupling coefficient is as close to a real number as possible (e. g., by adjusting the length of the coupling part). The following analysis applies to such a structure. In this case, oscillation frequency is equal to the average of those of the elements as shown eq. (18).

Under the above approximations, if $\kappa_{r(nk)} = 0$ except for $k = n \pm 1$ (nearest-neighbor coupled array), $\sin(\phi_k - \phi_n)$ in eq. (16) can be calculated analytically (see Appendix 1). For an array with an arbitrary coupling, however, analytical expression is not possible, and statistical analysis of the frequency variation becomes unpredictable.

Therefore, the following approximation is made. First, when the oscillation frequencies of all elements are the same, eq. (16) with $\kappa_{i(nk)} = 0$ results in $\sin(\phi_k - \phi_n) = 0$, and thus, the phase difference $\phi_k - \phi_n$ is either 0 (in-phase) or $\pm\pi$ (opposite phase) depending on the oscillating supermode. The oscillating supermode is determined from the coefficient matrix in eq. (15) as mentioned above [33] or from the matrix of $\kappa_{r(nk)}$ as shown in Appendix 4.

When the oscillation frequency of each element varies, $\phi_k - \phi_n$ deviates from 0 or $\pm\pi$. Expressing the phase difference under the frequency variation as $\phi_n = \phi_n^{(0)} + \phi_n^{(1)}$, where $\phi_n^{(0)} = 0$ or $\pm\pi$, the following approximation is obtained.

$$\sin(\phi_k - \phi_n) \simeq \sin\left(\phi_k^{(0)} - \phi_n^{(0)}\right) + \cos\left(\phi_k^{(0)} - \phi_n^{(0)}\right) \cdot \left(\phi_k^{(1)} - \phi_n^{(1)}\right) = \pm\left(\phi_k^{(1)} - \phi_n^{(1)}\right), \quad (19)$$

where the signs $\pm$ on the right-most side correspond to $\phi_k^{(0)} - \phi_n^{(0)} = 0$ for $+$ and $\pm\pi$ for $-$. The same meaning applies when $\pm$ appears in the equations below except for those in Fig. 5. Substituting eq. (19) in eq. (16) gives the following equation.

$$\frac{\omega_n - \omega}{\omega_0} + \sum_{k=0}^{N-1}\left(\pm\kappa_{r(nk)}\right)\left(\phi_k^{(1)} - \phi_n^{(1)}\right) = 0 \qquad (n = 0,1,\cdots N-1). \qquad (20)$$

These equations are linear simultaneous equations for $\phi_n^{(1)}$ $(n = 0,1,\cdots N-1)$ and analytically solvable. The difference between the exact solution $\sin(\phi_k - \phi_n)$ of eq. (16) and the approximate solution $\pm\left(\phi_k^{(1)} - \phi_n^{(1)}\right)$ of eq. (20) is about 10% for the same frequency variation, even near the limit of mutual locking where these solutions are close to 1 (see Appendix 2). Furthermore, when only adjacent elements are coupled as described above, the approximate solution completely coincides with the exact solution.

As the phase reference is arbitrary in eq. (20), $\phi_0^{(0)} = \phi_0^{(1)} = 0$ by setting $\phi_0$ to the reference. $\omega$ in eq. (20) is given by the first term of eq. (18). The unknowns in eq. (20) are therefore $\phi_n^{(1)}$ $(n = 1,2,\cdots,N-1)$, and the number of unknowns is $N-1$. As a result, $N-1$ equations of eq. (20), excluding that for $n = 0$, are used in the analysis.

The solution of eq. (20) is obtained as (see Appendix 3)

$$\sin(\phi_k - \phi_n) \simeq \pm\left(\phi_k^{(1)} - \phi_n^{(1)}\right) = \pm\frac{1}{\omega_0\det(A)}\left\{\frac{\det(B^{(kn)})}{N}\sum_{m=0}^{N-1}\omega_m - \sum_{m=1}^{N-1}\left[\omega_m\,(-1)^{k+m}\Delta_{mk}^{(nk)}\right]\right\}, \tag{21}$$

where $A$ and $B^{(nk)}$ are $(N-1)\times(N-1)$ matrices with the following elements,

$$A_{ij} = \begin{cases} \pm\kappa_{r(ij)} & \text{for off diagonal } (i \neq j \,;\, i = 1,2,\ldots,N-1), \\[2mm] -\sum_{m=0}^{N-1}\left(\pm\kappa_{r(im)}\right) & \text{for diagonal } (i = j \,;\, i = 1,2,\ldots,N-1), \end{cases} \tag{22}$$

$$B^{(kn)}_{ij} = \begin{cases} 1 & \text{for the k} - \text{th column } (j = k \,;\, i = 1,2,\ldots,N-1), \\ A_{in} + A_{ik} & \text{for the n} - \text{th column } (j = n \,;\, i = 1,2,\ldots,N-1), \\ A_{ij} & \text{for others } (j \neq k,n \,;\, i,j = 1,2,\ldots,N-1), \end{cases} \tag{23}$$

and $\Delta_{mk}^{(kn)}$ in eq. (21) is the determinant of the adjugate matrix of $B^{(kn)}$, i.e., the determinant of the $(N-2) \times (N-2)$ matrix removed the $m$-th column and $k$-th row from $B^{(kn)}$. Equation (21) is the solution for $n, k = 1, 2, \cdots, N-1$, but it can express the solution for $n = 0$ by replacing $B^{(kn)}$ with $A'$, where $A'$ is given by eq. (22) except that all the elements on the $k$-th column are 1. Equation (21) can also express the solution for $k = 0$ by swapping $k$ and $n$, changing the sign, and replacing $B^{(kn)}$ with $A'$.

If $B^{(kn)}$ is a regular matrix, its inverse matrix exists, and $\left(B^{(kn)}\right)_{km}^{-1} = (-1)^{k+m} \Delta_{mk}^{(kn)} / \det\left(B^{(kn)}\right)$ holds. Thus, eq. (21) is reduced to

$$\sin(\phi_k - \phi_n) \simeq \pm\left(\phi_k^{(1)} - \phi_n^{(1)}\right) = \pm\frac{\det(B^{(kn)})}{\omega_0 \det(A)} \left\{ \frac{1}{N}\sum_{m=0}^{N-1} \omega_m - \sum_{m=1}^{N-1}\left[\omega_m \left(B^{(kn)}\right)_{km}^{-1}\right] \right\}. \qquad (24)$$

$B^{(kn)}$ is often not regular when the coupling is complex, and it is necessary to use eq. (21) in that case.

Let the right-most side of eq. (21) or (24) be $\Omega_s$ as in the analysis of a two-element array. The condition for mutual locking between the elements $n$ and $k$ can be written as $-1 \leq \Omega_s \leq 1$. Since stable operation of array requires that all the coupled elements satisfy this condition, this condition must be examined for the elements with the largest $|\Omega_s|$.

If the variations in the oscillation angular frequencies ($\omega_m$'s in eq. (21) or (24)) are all independent and follow the same Gaussian distribution with the mean value $\langle\omega\rangle$ and variance $\langle\Delta\omega^2\rangle$, the variation in $\Omega_s$ also follows a Gaussian distribution, as in the case of two elements, and its mean value and variance are given by (see Appendix 3)

$$\langle\Omega_s\rangle = 0, \qquad (25)$$

$$\langle\Delta\Omega_s^2\rangle = \frac{1}{(\det(A))^2}\left[\sum_{m=1}^{N-1}\left(\Delta_{mk}^{(kn)}\right)^2 - \frac{1}{N}\left(\det\left(B^{(kn)}\right)\right)^2\right]\frac{\langle\Delta\omega^2\rangle}{\omega_0^2} \quad \text{for regular } B^{(kn)} \qquad (26)$$

$$= \left(\frac{\det\left(B^{(kn)}\right)}{\det(A)}\right)^2\left[\sum_{m=1}^{N-1}\left(\left(B^{(kn)}\right)_{km}^{-1}\right)^2 - \frac{1}{N}\right]\frac{\langle\Delta\omega^2\rangle}{\omega_0^2} \quad \text{for irregular } B^{(kn)}. \qquad (27)$$

As in the case of two elements, the probability of mutual locking is given by $p = \text{Erf}\left(1/\sqrt{2\langle\Delta\Omega_s^2\rangle}\right)$, and thus, the relative frequency variation tolerance for mutual locking at a specified $p$ is written as follows from eqs. (26) and (27) using the coefficient $c(p)$ in Fig. 3:

$$\frac{\sqrt{\langle\Delta f^2\rangle}}{f_0} = c(p)\,|\det(A)|\left/\sqrt{\sum_{m=1}^{N-1}\left(\Delta_{mk}^{(kn)}\right)^2 - \frac{1}{N}(\det(B^{(kn)}))^2}\right. \quad \text{for regular } B^{(kn)} \qquad (28)$$

$$= c(p)\left|\frac{\det(A)}{\det(B^{(kn)})}\right| \left/ \sqrt{\sum_{m=1}^{N-1}((B^{(kn)})_{km}^{-1})^2 - \frac{1}{N}}\right. \quad \text{for irregular } B^{(kn)}, \qquad (29)$$

where the combination $n$ and $k$ of the elements in eqs. (26) or (27) is chosen so that $\sqrt{\langle \Delta f^2 \rangle}/f_0$ has the smallest value.

Equation (28) or (29) can be applied to any coupling configuration, i.e., any $\kappa_{r(ij)}$, except for the special case where the couplings are divided into two or more independent groups. In this case, $\kappa_{r(ij)}$ can still be defined, but the array as a whole will not shows mutual locking, and neither eq. (28) nor (29) has a solution.

Below, some examples of calculations for eqs. (28) and (29) are presented. As a first example, the frequency variation tolerance for $N$-element 1D coupled arrays is calculated. Arrays in which each element is coupled up to its nearest neighbors or its second-nearest neighbors are considered. Figure 5 shows these coupling configurations. The absolute value of $\kappa_{r(ij)}$ is assumed the same for all coupled elements, $|\kappa|$, and its sign is selected for each element so that the desired supermode is obtained (see Appendix 4). (Note that the sign of $|\kappa|$ in Fig. 5 does not mean $\pm$ in eq. (19)). Although a slot antenna array is shown in Fig. 5 as an example, the structures of the element and coupling are not limited to this [27, 33-35].

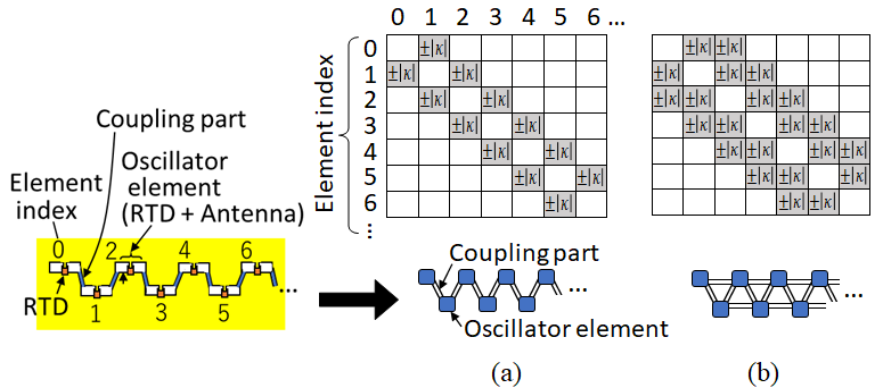

**Fig. 5** Coupling configurations and matrices of coupling coefficient ($\kappa_{r(ij)}$) of 1D coupled arrays for which the frequency variation tolerance was calculated. The coupling configuration is shown using a slot antenna array as an example. (a) A nearest-neighbor coupled array, and (b) an array with couplings up to the second-nearest neighbors. The absolute values of the coupling coefficients are all the same, $|\kappa|$, and the sign of each $|\kappa|$ is selected so that the desired supermode is obtained. For example, if all are positive, all elements oscillate in phase, i.e., in the fundamental supermode. Blanks in the matrices indicate 0 (no couplings).

In Fig. 5(a) and (b), if the coupling coefficients between the coupled elements are all positive, all elements oscillate in phase (fundamental supermode oscillation). For the nearest-neighbor couplings in Fig. 5(a), if the coupling coefficients between the coupled elements are all negative, the highest-order supermode oscillates, where adjacent elements are in anti-phase. The calculation result in this case is the same as that for the fundamental supermode oscillation, as seen from equation (20).

In Fig. 5(b), where each element is coupled to the nearest and second nearest elements, if the coupling coefficient for the nearest neighbors is negative and that for the second nearest neighbors is positive, the highest supermode oscillates. As with the case for the nearest-neighbor couplings described above, the result will be the same as that of the fundamental supermode.

The calculation results for 1D arrays are shown in curves (a) and (b) in Fig. 6, which show the frequency variation tolerance required for mutual locking to occur with 95% probability, i.e., if the frequency variation of the elements is kept within the calculated values in device fabrication, the locking will occur with 95% probability. The curve (a) is the result for the nearest-neighbor coupled array, and (b) is that for the array with couplings up to the second-nearest neighbors. Curves (c) and (d) are the results for 2D arrays discussed later.

From the curves (a) and (b) in Fig. 6, the frequency variation tolerance changes very slightly for small element number $N$, but decreases inversely proportional to $\sqrt{N}$ for large $N$. Furthermore, due to the larger number of couplings, the frequency variation tolerance in (b) is larger than that in (a). The result in (a) exactly agrees with the analytical results (see Appendix 1). The frequency variation tolerance is the smallest between the elements near the center of the array for (a) and (b).

As a numerical example, when $\mathrm{Re}(y_{ij}) \sim 2$ mS [33], $C_0 \sim 10$ fF, and $f_0 = \omega_0/2\pi \sim 1$ THz as in the case of a two-element array, $|\kappa| \sim 0.02$ is obtained from eq. (17), and the vertical scale 1 in Fig. 6 corresponds to $\sqrt{\langle \Delta f^2 \rangle}/f_0 \sim 0.02$ ($\langle \Delta f^2 \rangle \sim 20$ GHz). A strong coupling (a large $|\kappa|$) gives a large frequency variation tolerance.

As a second example, the frequency variation tolerance of 2D arrays is calculated. The 2D array structure schematically shown in Fig. 7(a) can be constructed by slot antennas with couplings to other elements from both ends and the center of the slot [35]. Patch antennas can also be used by connecting with each other through transmission lines as in ref. [27]. However, the structure in [27] contains two RTDs in one patch antenna. The calculation results for this structure will be discussed later.

Figure 7(b) shows the dependence of the frequency variation tolerance on element number in the longitudinal direction $N_V$, with the element number in the lateral direction $N_H$, fixed. The coupling coefficient $\kappa$ is assumed the same for all elements. When $\kappa > 0$, the fundamental supermode oscillates, in which all elements oscillate in phase, while when $\kappa < 0$, the highest-order supermode oscillates, in which adjacent RTDs oscillate in opposite phases. The results are the same for both modes.

The frequency variation tolerance increases with increasing $N_V$ from a 1D array ($N_V = 1$), and reaches its maximum when the 2D array becomes square ($N_V = N_H$). As $N_V$ increases further, the frequency variation tolerance decreases. This decrease is smaller than $\propto 1/\sqrt{N}$ of 1D arrays. The frequency variation tolerance of 2D square arrays decreases as the element number increases, but the decrease is relatively gradual.

The frequency variation tolerance of the nearest-neighbor coupled 2D square arrays (i.e., the peak point of each curve in Fig. 7) is shown in Fig. 6 by curve (c). In the 2D square arrays, the decrease in frequency variation tolerance with increasing element number $N$ is smaller than in 1D arrays, indicating that 2D arrays are advantageous for multi-element coupled arrays. In the 2D square arrays, the frequency variation tolerance is the smallest between the elements near the center of the rows at square sides.

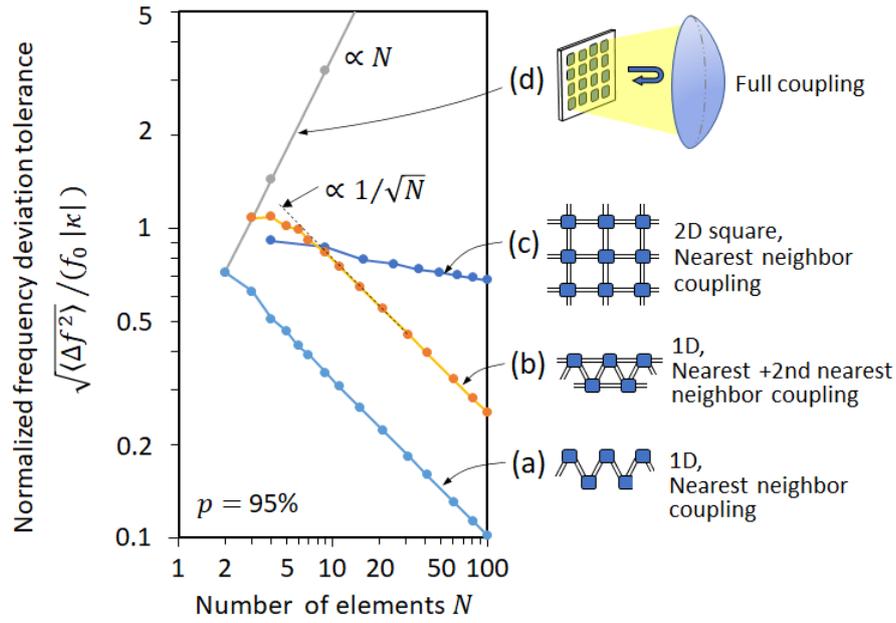

**Fig. 6** Dependence of the frequency variation tolerance on element number $N$ for the probability of mutual locking $p = 95\%$. (a) nearest-neighbor coupled 1D array, (b) 1D array with couplings up to second-nearest neighbors, (c) nearest-neighbor coupled 2D square array, and (d) array with all elements coupled. Each curve represents the case where the supermode is the fundamental, in which all elements oscillate in phase, or the highest-order one, in which adjacent RTDs oscillate in opposite phases. The results are the same for both modes. The curves are slightly less smooth because of the dependence on whether $N$ is even or odd. This situation is shown for the curve (a) by eqs. (A2-7) and (A2-8) in Appendix 1.

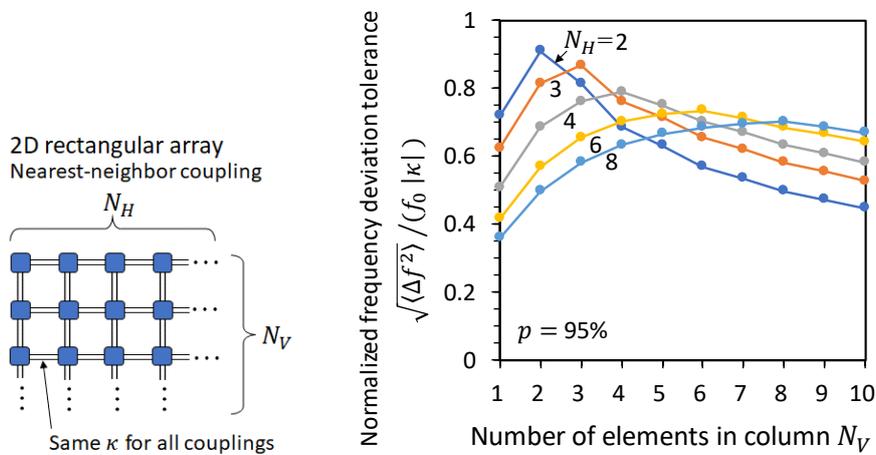

**Fig. 7** Frequency variation tolerance for 2D arrays for the probability of mutual locking $p = 95\%$. (a) Schematic array configuration of a nearest-neighbor coupled 2D array, and (b) dependence of frequency variation tolerance on longitudinal element number $N_V$ with the lateral element number $N_H$ fixed. The coupling coefficient $\kappa$ is assumed the same for all elements. When $\kappa > 0$, the fundamental supermode oscillates in which all elements oscillate in phase, while when $\kappa < 0$, the highest-order supermode oscillates, in which adjacent RTDs oscillate in opposite phases. The results are the same for both modes.

The calculation result for fully coupled arrays (arrays in which each element is coupled to all other elements) is also shown by curve (d) in Fig. 6. The frequency variation tolerance increases in proportion to $N$, because the number of couplings increases at the same rate as $N$. In an actual fully coupled array, the elements can be coupled either by feeding the combined total output equally back to all elements, as shown in the inset of Fig. 6, or by placing all elements within a single cavity. In these cases, although the frequency variation tolerance normalized by $|\kappa|$ increases with $N$ as shown in Fig. 6, the frequency variation tolerance does not necessarily increase with $N$ in practice because $|\kappa|$ decreases with $N$ due to the division of coupled power.

As a final calculation example, Figure 8 shows the calculation results for 2D square arrays in which two RTDs are integrated into one patch antenna as one element, which corresponds to the configuration in ref. [27]. The coupling coefficients used in the calculation are $\kappa_2 = -2\kappa_1$ and $\kappa_3 = -22\kappa_1$ (see Appendix 5). Due to this sign relationship of the coupling coefficients, the oscillation voltages of adjacent RTDs are in opposite phase, as shown in Fig. 8(a). This phase relationship results in in-phase radiation above the patch array.

As shown in Fig. 8(b), the frequency variation tolerance decreases as the element number increases. The rate of decrease is large when the element number is small, but the decrease is gradual when the element number is large. Figure 8(b) also shows the results for the 2D square arrays. The curve for the 1-patch 2-RTD configuration is similar to that of the 2D square array. This is because the coupling within the patch is so strong, i.e., $|\kappa_3|$ is so large, that two RTDs in the same patch can be regarded as a single oscillator. The frequency variation tolerance for the 1-patch 2-RTD configuration is larger than that for the square array because $|\kappa_2| > |\kappa_1|$ in the present calculation.

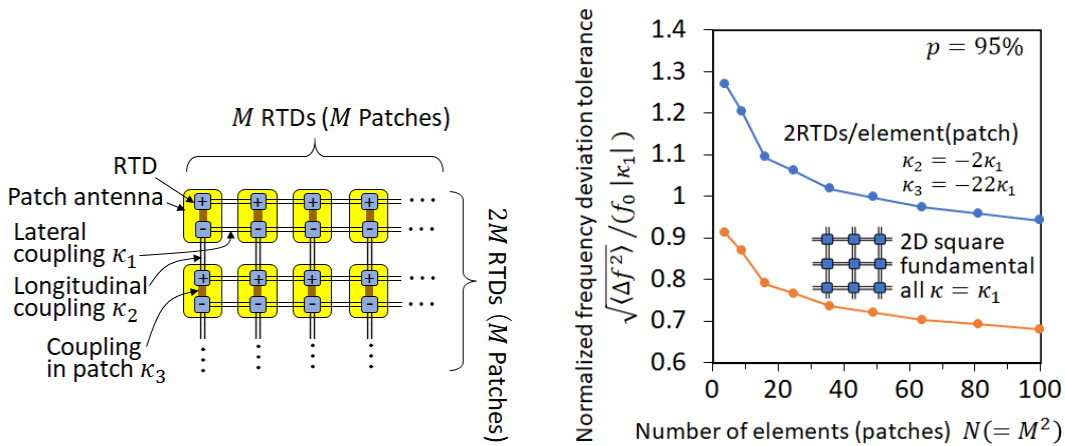

**Fig. 8** Frequency variation tolerance in 2D square arrays where one element is composed of two RTDs and a patch antenna. Probability of mutual locking $p = 95\%$. (a) Array configuration, where $+$ and $-$ indicate the phase relationship of the voltage amplitude of each RTD. (b) Dependence of the frequency variation tolerance on the element number (patch number). Results for the 2D square arrays are also shown.

Comparison with the experimental results in ref. [27] is possible if the experimental value of $\kappa_1$ and frequency variation conditions are given. Considering stable mutual locking has been obtained with 36 elements in ref. [27], the frequency variation is likely to be sufficiently smaller than the tolerance shown in Fig. 8(b). Since the decrease in frequency variation tolerance with element number is calculated to be small, it seems quite possible to increase the element number while maintaining mutual locking in the experiment. However, comprehensive consideration is required, such as the increase in heat generation with the increase in element number and the design of circuits to prevent parasitic oscillation.

## 4. Summary

An approximate theoretical formula was derived on the frequency variation tolerance for mutual locking between elements in an array oscillator. This formula was applied to 1D and 2D arrays, and the element-number dependence of the frequency variation tolerance was shown. In 1D arrays, the frequency variation tolerance decreases inversely proportional to the square root of the element number. In 2D arrays, the frequency variation tolerance has a maximum in square configuration, and the maximum value decreases with increasing element number, but the dependence on element number is smaller than that of 1D arrays. Calculation was also performed for 2D square arrays in which each element is composed of two RTDs and a patch antenna, and the element-number dependence of the frequency variation tolerance is shown to be nearly identical to that of nearest-neighbor coupled square arrays.

The derived approximate formula for the frequency variation tolerance can be applied to any coupling topology, and can therefore be used as a tool for exploring coupling topology that achieves mutual locking even with large frequency variations.

Acknowledgement

The authors would like to thank Honorary Prof. Y. Suematsu for continuous encouragement. This study was supported by Grant-in-Aid for Scientific Research (24H00031) from JSPS, CREST (JPMJCR21C4) from JST, X-NICS (JPJ011438) and ARIM (JPMXP1225IT0018, JPMXP1225IT0019) from MEXT, the Innovative Science & Technology Initiative for Security (JPJ013268) from ATLA, and ROHM Co., Ltd.

**Appendix 1** Exact Analysis of Nearest-Neighbor Coupled Arrays with Real-Number Coupling Coefficients

If each element in an array couples only with its adjacent elements (i.e., $\kappa_{r(nk)} = 0$ except for $k = n \pm 1$ in Eq. (16)), $\sin(\phi_k - \phi_n)$ in Eq. (16) can be expressed analytically.

For simplicity, the following discussion focuses on the case where all non-zero coupling coefficients have the same value ($\kappa_{r(nk)} = \kappa$) as shown in Fig. 5(a). Even when the coupling coefficients are not all the same, an analytical expression is possible, and the statistical process for frequency variation can be performed without approximating eq. (16) to the linear equation (eq. (20)). The same approximations as in the main text are used here except for the linear approximation. In this case, eq (16) becomes:

$$\left.\begin{array}{l} \dfrac{\omega_0 - \omega}{\omega_0} + \kappa \sin(\phi_1 - \phi_0) = 0 \,, \\[3mm] \dfrac{\omega_n - \omega}{\omega_0} + \kappa \sin(\phi_{n+1} - \phi_n) + \kappa \sin(\phi_{n-1} - \phi_n) = 0 \,, \quad (n = 1, \cdots N-2) \\[3mm] \dfrac{\omega_{N-1} - \omega}{\omega_0} + \kappa \sin(\phi_{N-2} - \phi_{N-1}) = 0 \,. \end{array}\right\} \qquad (A1-1)$$

Equation (A1-1) can be regarded as a linear simultaneous equation with respect to $\sin(\phi_{n+1} - \phi_n)$ instead of $\phi_k^{(1)} - \phi_n^{(1)}$. This is the reason why analytical expression is possible for the nearest-neighbor coupled arrays and the linear approximation gives the exact solution. Equation (18) also holds in this case, neglecting the second term. Equation (A1-1) can be solved analytically by solving the equations from the one with smaller $n$ in order, and the following solution is obtained.

$$\sin(\phi_{n+1} - \phi_n) = \frac{1}{\omega_0 \kappa} \left[ \frac{n+1}{N} \sum_{k=n+1}^{N-1} \omega_k - \left(1 - \frac{n+1}{N}\right) \sum_{k=0}^{n} \omega_k \right]. \; (n = 0, \cdots N-1) \qquad (A1-2)$$

Let the right-hand side be $\Omega_s$. Its mean value and squared variance are given as follows similarly to the main text and Appendix 3.

$$\langle \Omega_s \rangle = 0 \,, \qquad\qquad\qquad\qquad\qquad\qquad\qquad\qquad\qquad (A1-3)$$

$$\langle \Delta\Omega_s^2 \rangle = \left[ \left(\frac{n+1}{N}\right)^2 (N-n-1) + \left(1 - \frac{n+1}{N}\right)^2 (n+1) \right] \frac{\langle \Delta\omega^2 \rangle}{(\omega_0 \kappa)^2}$$

$$= \frac{(N-n-1)(n+1)}{N} \frac{\langle \Delta\omega^2 \rangle}{(\omega_0 \kappa)^2} \,. \qquad\qquad\qquad (A1-4)$$

$\langle \Delta\Omega_s^2 \rangle$ in eq. (A1-4) varies with $n$, and its maximum value is given by

$$\langle \Delta\Omega_s^2 \rangle_{max} = \frac{N}{4} \frac{\langle \Delta\omega^2 \rangle}{(\omega_0 \kappa)^2} \quad \text{at } n = \frac{N}{2} - 1 \text{ for even } N \qquad (A1-5)$$

and

$$\langle \Delta\Omega_s^2 \rangle_{max} = \frac{N^2-1}{4N} \frac{\langle \Delta\omega^2 \rangle}{(\omega_0 \kappa)^2} \quad \text{at } n = \frac{N-1}{2} - 1 \text{ and } \frac{N-1}{2} \text{ for odd } N \,. \qquad (A1-6)$$

$\sqrt{\langle\Delta\Omega_z^2\rangle_{max}}$ is written as $c(p)$ (Fig. 3) after the same process as that in the main text. The frequency variation tolerance to ensure the probability $p$ for mutual locking can be calculated from eqs. (A1-5) and (A1-6) as

$$\frac{\sqrt{\langle\Delta f^2\rangle}}{f_0} = 2c(p)\sqrt{\frac{1}{N}}\,|\kappa| \qquad \text{for even } N\,, \qquad (A1-7)$$

and

$$\frac{\sqrt{\langle\Delta f^2\rangle}}{f_0} = 2c(p)\sqrt{\frac{N}{N^2-1}}\,|\kappa| \quad \text{for odd } N. \qquad (A1-8)$$

Although the frequency variation tolerance is different for even and odd $N$, in both cases it is inversely proportional to $\sqrt{N}$ for large $N$. This situation is shown in curve (a) of Fig. 6, which shows the result approximated by a linear simultaneous equation, but it exactly agrees with the analytical solution presented here.

**Appendix 2** Accuracy of Linear Approximation

Comparison is shown here between the results with and without the approximation $\sin(\phi_k - \phi_n) \simeq \pm(\phi_k^{(1)} - \phi_n^{(1)})$ in eq. (19). The equation without this approximation is

$$\frac{\omega_n - \omega}{\omega_0} + \sum_{k=0}^{N-1}\kappa_{r(nk)}\,\sin(\phi_k - \phi_n) = 0 \qquad (n = 0,1,\cdots N-1)\,, \qquad (A2-1)$$

and that with the approximation is

$$\frac{\omega_n - \omega}{\omega_0} + \sum_{k=0}^{N-1}(\pm\kappa_{r(nk)})(\phi_k^{(1)} - \phi_n^{(1)}) = 0 \qquad (n = 0,1,\cdots N-1)\,. \qquad (A2-2)$$

These two equations are numerically solved and the results are compared. As an example, a 1D array, in which each element is coupled up to its nearest and second nearest neighbors, is considered (Fig. 5(b)). All non-zero coupling coefficients are assumed equal ($\kappa_{r(nk)} = \kappa$).

Figure A-1 shows the calculation results for element number $N = 7$ and $\kappa = 0.02$. Figure A-1(a) shows the distribution of frequency variation assumed to each element, where $\Omega_i = \omega_i/\langle\omega\rangle \simeq \omega_i/\omega_0$, and the frequencies are linearly distributed. The difference in $\Omega_i$ between the center and edge elements is $\pm\delta$, for which the mean and standard deviation are given by $\langle\Omega\rangle = 1$ and $\Delta\Omega_{dev} = \sqrt{\sum_{i=0}^{N-1}(\langle\Omega\rangle - \Omega_i)^2/N} \simeq 0.317\delta$, respectively. (Note that $\Delta\Omega_{dev}$ has a different meaning from the frequency variation in device fabrication discussed in the main text. $\Delta\Omega_{dev}$ represents the frequency variation among the $N$ elements in the present calculation example.) Figure A-1(b) plots the calculation results as a function of $\Delta\Omega_{dev}$ with the variation of $\delta$. The Solid and dashed lines show the results of $\sin(\phi_k - \phi_n)$ and $\phi_k^{(1)} - \phi_n^{(1)}$ for each coupled element, respectively.

Mutual Locking is limited to the region of $\Delta\Omega_{dev}$ in which the calculation results ($\sin(\phi_k - \phi_n)$ and $\phi_k^{(1)} - \phi_n^{(1)}$) lie between -1 and 1. $\sin(\phi_k - \phi_n)$ in the exact calculations natually does not show a value greater than 1, while $\phi_k^{(1)} - \phi_n^{(1)}$ in the approximate calculation increases linearly beyond 1, as shown in Fig. A-1(b). The accuracy of the approximate calculation can be seen by comparing the values of $\Delta\Omega_{dev}$ at the calculation results = 1. In the example of Fig. A-1(b), the values of $\sin(\phi_k - \phi_n)$ and $\phi_k^{(1)} - \phi_n^{(1)}$ between elements 2 and 4 are largest, and the difference in $\Delta\Omega_{dev}$ where they become 1 is approximately 12%. Similar results are obtained for other frequency distributions, with the difference in $\Delta\Omega_{dev}$ being around 10% in all cases. In Fig. A-1(b), after $\sin(\phi_k - \phi_n)$ reaches 1 with increasing $\Delta\Omega_{dev}$, there is a part where it decreases slightly as $\Delta\Omega_{dev}$ further increases, but mutual locking is unstable in this part.

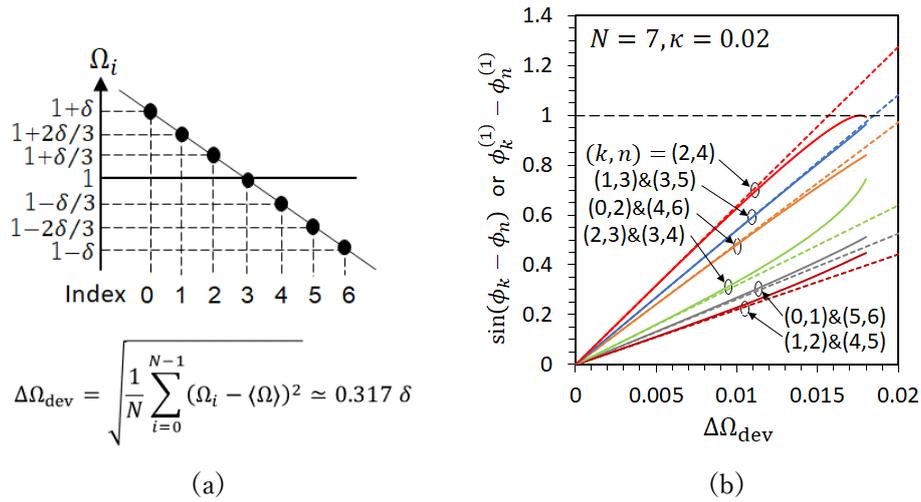

(a)  (b)

**Fig. A-1** Comparison of results between exact and approximated calculations. A 1D array in which each element is coupled up to its nearest and second nearest neighbors. (a) Assumed frequency variation between elements. Frequency is normalized by $\langle\omega\rangle$. (b) Calculation results of $\sin(\phi_k - \phi_n)$ (solid line) and $\phi_k^{(1)} - \phi_n^{(1)}$ (dotted line) for each coupled element as a function of $\Delta\Omega_{dev}$ with $\delta$ varied.

**Appendix 3** Derivation of eqs. (21)-(27)

Equation (20) can be rewritten as follows for $n = 1, 2, \cdots, N-1$, by discarding the equation for $n = 0$, as explained in the main text:

$$\begin{bmatrix} -\sum_{m=0}^{N-1}(\pm\kappa_{r(1m)}) & \pm\kappa_{r(12)} & \cdots & \pm\kappa_{r(1,N-1)} \\ \pm\kappa_{r(21)} & -\sum_{m=0}^{N-1}(\pm\kappa_{r(2m)}) & \cdots & \kappa_{r(2,N-1)} \\ \vdots & & \ddots & \vdots \\ \pm\kappa_{r(N-1,1)} & \pm\kappa_{r(N-1,2)} & \cdots & -\sum_{m=0}^{N-1}(\pm\kappa_{r(N-1,m)}) \end{bmatrix} \begin{bmatrix} \phi_1^{(1)} \\ \phi_2^{(1)} \\ \vdots \\ \phi_{N-1}^{(1)} \end{bmatrix} = \begin{bmatrix} (\omega - \omega_1)/\omega_0 \\ (\omega - \omega_2)/\omega_0 \\ \vdots \\ (\omega - \omega_{N-1})/\omega_0 \end{bmatrix}.$$

$(A3-1)$

The solution $\phi_k^{(1)}$ $(k = 1, 2, \cdots, N-1)$ of this equation is obtained as

$$\phi_k^{(1)} = \frac{\det(C_k)}{\det(A)} \ , \tag{A3-2}$$

where $A$ is a matrix composed of the coefficients of $\phi_k^{(1)}$ $(k = 1, 2, \cdots, N-1)$ in eq. (A3-1) and is given in eq. (22). $C_k$ is a matrix in which the element on the $m$-th row and $k$-th column of $A$ is replaced with $(\omega - \omega_m)/\omega_0$ $(m = 1, 2, \cdots, N-1)$, that is,

$$k\text{-th column}$$
$$\downarrow$$

$$C_k = \begin{bmatrix} -\sum_{m=0}^{N-1}(\pm\kappa_{r(1m)}) & \pm\kappa_{r(12)} & \cdots & (\omega - \omega_1)/\omega_0 & \cdots & \pm\kappa_{r(1,N-1)} \\ & & & \vdots & & \vdots \\ \pm\kappa_{r(21)} & -\sum_{m=0}^{N-1}(\pm\kappa_{r(2m)}) & \cdots & (\omega - \omega_2)/\omega_0 & \cdots & \pm\kappa_{r(2,N-1)} \\ \vdots & \vdots & & \vdots & & \vdots \\ \pm\kappa_{r(N-1,1)} & \pm\kappa_{r(N-1,2)} & \cdots & (\omega - \omega_{N-1})/\omega_0 & \cdots & -\sum_{m=0}^{N-1}(\pm\kappa_{r(N-1,m)}) \end{bmatrix} .$$

$$\tag{A3-3}$$

$\phi_n^{(1)}$ is also obtained similarly to $\phi_k^{(1)}$, and

$$\phi_k^{(1)} - \phi_n^{(1)} = \frac{\det(C_k) - \det(C_n)}{\det(A)} = \frac{\det(C_k) + \det(C_{n \leftrightarrow k})}{\det(A)}, \tag{A3-4}$$

where $C_{n \leftrightarrow k}$ is the matrix obtained by swapping the $n$-th and $k$-th columns of $C_n$ in eq. (A3-3). $C_k$ and $C_{n \leftrightarrow k}$ have the same matrix elements except for those on the $n$-th column. The $n$-th columns of $C_k$ and $C_{n \leftrightarrow k}$ are equal to the $n$-th and $k$-th columns of $A$, respectively. Therefore, $\det(C_k) + \det(C_{n \leftrightarrow k})$ is equal to the determinant of $C_k$ replaced the $n$-th column with the sum of the $n$-th and $k$-th columns of $A$. Furthermore, since the $k$-th column of this determinant is $(\omega - \omega_m)/\omega_0$ ($m$ = row number; $1, 2, \cdots, N-1$), this determinant can be decomposed into two determinants whose $k$-th columns are $\omega/\omega_0$ and $-\omega_m/\omega_0$. As a result,

$$\phi_k^{(1)} - \phi_n^{(1)} = \frac{\omega \det(B^{(kn)}) - \det(D^{(kn)})}{\omega_0 \det(A)} \ , \tag{A3-5}$$

where $B^{(kn)}$ is given by eq. (23) in the main text, and $D^{(kn)}$ is given by $B^{(kn)}$ replaced the $k$-th column ($= 1$) with $\omega_m$. Substituting the first term on the right-hand side of eq. (18) for $\omega$ in eq. (A3-5) (only the first term is used because of $\kappa_{i(nk)} = 0$), and expanding $D^{(kn)}$ with the product between the determinant of the adjugate matrix $\Delta_{mk}^{(kn)}$ and the elements in the $k$-th column, eq. (20) is obtained. If $B^{(kn)}$ is regular, equation (20) is further reduced to eq. (24).

Equations (21) and (A3-5) are expressions for $k, n = 1, 2, \cdots, N-1$, and cannot be applied when $\phi_k^{(1)}$ or $\phi_n^{(1)}$ is $\phi_0^{(1)} (= 0)$. $\det(C_k)$ must be set to 0 when $\phi_k^{(1)} = \phi_0^{(1)}$, and $\det(C_{n \leftrightarrow k})$ must be set to 0 when $\phi_n^{(1)} = \phi_0^{(1)}$ in $\det(C_k) + \det(C_{n \leftrightarrow k})$ in eq. (A-3-4). These operations result in the expressions of the solutions including $\phi_0^{(1)}$ by the modification of eq. (21), as described in the main text.

To derive eqs (25) – (27), let the rightmost side of eq. (21) be $\Omega_s$. If the variation of $\omega_m$ of each element is an independent Gaussian distribution, the variation of $\Omega_s$ is also a Gaussian distribution due to the reproductive property of Gaussian distributions, as $\Omega_s$ is a linear combination of $\omega_m$. The influence of the variation of $\omega_0$ in the denominator of eqs. (A3-5) and (21) is neglected as it is small compared to the total variation of $\omega_m$'s in the numerator.

From eq. (21),

$$\Omega_s = \sum_{m=0}^{N-1} a_m \, \omega_m \;, \tag{A3-6}$$

where

$$a_m = \begin{cases} \dfrac{1}{\omega_0 \det(A)} \dfrac{\det(B^{(kn)})}{N} & (m = 0) \;, \\[4mm] \dfrac{1}{\omega_0 \det(A)} \left( \dfrac{\det(B^{(kn)})}{N} - (-1)^{k+m} \Delta_{mk}^{(nk)} \right) & (m = 1, 2, \cdots, N-1) \;. \end{cases} \tag{A3-7}$$

From the reproductive property of Gaussian distributions,

$$\langle \Omega_s \rangle = \sum_{m=0}^{N-1} a_m \langle \omega_m \rangle = \langle \omega \rangle \sum_{m=0}^{N-1} a_m$$
$$= \frac{\langle \omega \rangle}{\omega_0 \det(A)} \left( \det(B^{(kn)}) - \sum_{m=1}^{N-1} (-1)^{k+m} \Delta_{mk}^{(nk)} \right) = 0 \;, \tag{A3-8}$$

$$\langle \Delta\Omega_s^2 \rangle = \sum_{m=0}^{N-1} a_m^2 \langle \Delta\omega_m^2 \rangle = \langle \Delta\omega^2 \rangle \sum_{m=0}^{N-1} a_m^2$$
$$= \frac{\langle \Delta\omega^2 \rangle}{(\omega_0 \det(A))^2} \left[ \left( \frac{\det(B^{(kn)})}{N} \right)^2 + \sum_{m=1}^{N-1} \left( \frac{\det(B^{(kn)})}{N} - (-1)^{k+m} \Delta_{mk}^{(nk)} \right)^2 \right]$$
$$= \frac{\langle \Delta\omega^2 \rangle}{(\omega_0 \det(A))^2} \left[ \frac{(\det(B^{(kn)}))^2}{N} + \sum_{m=1}^{N-1} \left( \Delta_{mk}^{(nk)} \right)^2 - \frac{2 \, \det(B^{(kn)})}{N} \sum_{m=1}^{N-1} (-1)^{k+m} \Delta_{mk}^{(nk)} \right]$$
$$= \frac{\langle \Delta\omega^2 \rangle}{(\omega_0 \det(A))^2} \left[ \sum_{m=1}^{N-1} \left( \Delta_{mk}^{(nk)} \right)^2 - \frac{(\det(B^{(kn)}))^2}{N} \right] \;, \tag{A3-9}$$

In deriving eqs. (A3-8) and (A3-9), $\det(B^{(kn)})$ is expanded with the product between the element $(B^{(kn)})_{nk} (= 1)$ and the determinant of the adjugate matrix $\Delta_{mk}^{(nk)}$. From eqs. (A3-8) and (A3-9), eqs. (25) and (26) are obtained. If $B^{(kn)}$ is regular, eq. (26) is further reduced to eq. (27) in the same calculation as that to obtain eq. (24).

**Appendix 4** Relationship between Distribution of Coupling-Coefficient and Supermode

The array oscillates in the supermode for which the real part of the eigenvalue of the coefficient matrix in eq. (15) is the smallest and negative. Here, the relationship between this eigenvalue and the eigenvalue of the matrix composed of the real part of the coupling coefficient $\kappa_{r(nk)}$ ($n, k = 0, 1, \cdots, N-1$) is examined to discuss which supermode oscillates depending on the configuration of the coupling coefficients. The matrix considered here has off-diagonal elements expressed by eq. (17), and diagonal elements that are zero. Since this matrix is a symmetric matrix with real elements, its eigenvalues are real. The following relationship holds between the eigenvalue $\gamma_\kappa$ of this matrix and the eigenvalue $\gamma$ of the coefficient matrix in eq. (15):

$$\gamma_\kappa = -\frac{\omega_0 L}{2}\mathrm{Re}(\gamma) - \frac{\omega_0 L}{2}\left(G_{RTD} - G_L\right), \qquad (A4-1)$$

where $G_L$ is the real part of $y_{nn}$ indicated in Fig. 4. The difference in $y_{nn}$ between the array elements is neglected as it is small and does not affect the oscillation mode. Also, $v_n$ included in $G_{RTD}$ is set to 0, as in the main text. Each eigenvalue $\gamma_\kappa$ has the same eigenvector as the corresponding $\gamma$, and the distribution of $v_n$ of the supermode can be obtained from these eigenvectors.

From eq. (A4-1), the minimum $\mathrm{Re}(\gamma)$ corresponds to the maximum $\gamma_\kappa$, and thus, the supermode with the maximum $\gamma_\kappa$ oscillates. For example, in the matrix of Fig. 5(a), if all matrix elements are positive, all components of the eigenvector for the maximum $\gamma_\kappa$ have the same sign, which implies oscillation of the fundamental supermode. If all matrix elements are negative, the maximum $\gamma_\kappa$ is the same, but its eigenvector has alternating positive and negative components, i.e., oscillation of the highest order supermode. In the case of Fig. 5(b), if all matrix elements are positive, the fundamental supermode oscillates. If the matrix elements between the nearest-neighbor elements are negative and that between the second nearest-neighbor elements are positive, the highest-order supermode oscillates.

**Appendix 5**  Mutual Admittance and Coupling Coefficient of a Coupling Part with a Microstrip Line

The mutual admittance $y_{12}$ of a two-terminal pair network using a microstrip transmission line is calculated here. The length, width, and height from the ground plane are $\ell$, $W$, and $h$, respectively, as shown in Fig. A-2. The relative dielectric constant of the dielectric layer is $\varepsilon_r$. This structure is the coupling part between elements in the patch-antenna array in ref. [27], and its mutual admittance approximately corresponds to the off-diagonal elements of the $Y$ matrix in Figs. 1(b) and 4.

The admittance matrix of the two-terminal pair network in Fig. A-2 is given by

$$Y = Y_0 \begin{bmatrix} \coth(\gamma\ell) & -\operatorname{cosech}(\gamma\ell) \\ -\operatorname{cosech}(\gamma\ell) & \coth(\gamma\ell) \end{bmatrix}, \qquad (A5-1)$$

where $\gamma = \alpha + j\beta$ with $\alpha$ and $\beta$ being the loss coefficient and the propagation constant, respectively, and $Y_0$ is the characteristic admittance of the transmission line.

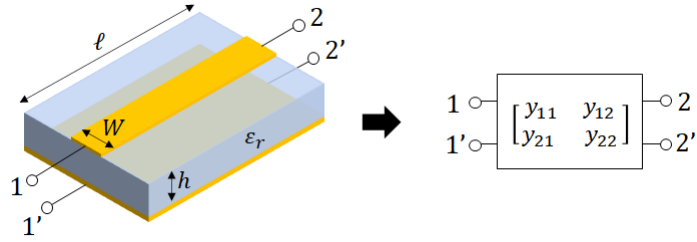

**Fig. A-2** Two-terminal pair network with microstrip transmission line and its $Y$ matrix.

The mutual admittance (off-diagonal elements of the admittance matrix) is written

$$y_{12} = -Y_0 \operatorname{cosech}(\gamma \ell) = \frac{-Y_0}{\sinh(\alpha \ell)\cos(\beta \ell) + j\sin(\beta \ell)\cosh(\alpha \ell)} . \quad (A5-2)$$

From equation (A5-2), the mutual admittance, and therefore the coupling coefficient in eq. (17), can be adjusted by the length $\ell$.

By choosing $\ell$ so that $\beta \ell = n\pi$ ($n = 1, 2, \cdots$), i.e., $\ell = n\,\lambda_g/2$ with $\lambda_g$ being the wavelength in the transmission line, $y_{12}$ is calculated as

$$y_{12} \simeq \pm \frac{Y_0}{\alpha \ell} \quad (\alpha \ell \ll 1), \quad (A5-3)$$

where $+$ and $-$ represent the cases for odd and even $n$, respectively.

When the transmission loss is small, $Y_0$ is approximately a real number. Equation (A5-3) is real in this case, and the coupling coefficient in eq. (17) is also real. In ref. [27], the lengths of the coupling parts with the coupling coefficients $\kappa_1$ and $\kappa_2$ in Fig. 8(a) are $\lambda_g$ and $\lambda_g/2$, respectively. Therefore, from equation (A5-3), $\kappa_1$ and $\kappa_2$ are approximately real numbers, and $\kappa_2 = -2\kappa_1$.

The patch antennas in Fig. 8(a) correspond to wide transmission lines, and their lengths are roughly half the wavelength. Therefore, the ratio of $\kappa_2$ to $\kappa_3$ is equal to the ratio of the characteristic admittance of the coupling transmission line to that of the patch antenna, as determined by eq. (A5-3). The characteristic impedance of the coupling transmission line ($1/Y_0$) is designed to be $100\,\Omega$ in ref. [27]. The characteristic admittance of the patch antenna is approximately given by $Y_0 \simeq (\sqrt{\varepsilon_r}/120\pi)\,(W/h)$ for $W \gg h$, and is calculated to be $0.11$ S with $W = 170\,\mu$m, $h = 7\,\mu$m, and $\varepsilon_r = 2.7$ (benzocyclobutene) [27]. Therefore, $\kappa_3 = 11\kappa_2 = -22\kappa_1$. Because of the strong coupling of the RTDs inside the patch antenna, $|\kappa_3| \gg |\kappa_1|$. Under this condition, the frequency variation tolerance is insensitive to the change in $\kappa_3$.